\newcounter{secnum}

\newcommand{\mysection}[1]{%
\vspace{1.25\baselineskip}
\stepcounter{section}
\stepcounter{secnum}
\centerline{\large\bf\thesecnum. #1}
\vspace{1pt}
}

\newcounter{subsecnum}[secnum]

\newbox\slashbox
\newdimen\slashwd
\newcommand{\slashed}[1]{%
\setbox\slashbox=\hbox{$#1$}%
\slashwd=\wd\slashbox%
\hbox to\slashwd{\hss/\hss}%
\llap{$#1$}}

\newcommand{\pslash}{\slashed{p}}

\documentstyle[12pt]{article}

\begin{document}
 
\begin{titlepage}
\begin{center}
{\bf Is the Standard Model Renormalizable ?} 
\end{center} 

\begin{center}
Hung Cheng\\ 
Department of Mathematics, Massachusetts Institute of Technology\\
Cambridge, MA 02139\\

\bigskip
and
\bigskip

S.P. Li\\
Institute of Physics, Academia Sinica, Nankang\\ 
Taipei, Taiwan, Republic of China\\ 
\end{center}

\end{titlepage}

\begin{abstract} 

In this paper, we study the renormalizability of the standard model in the
Landau gauge.  On the basis of the Ward-Takahashi identities, we derive 
exact expressions for the physical masses of $W$ and $Z$ as well as the
renormalized coupling constants in the theory. We show that it is
impossible to make all these renormalized quantities finite.
Thus the quantum theory of the standard model with the divergent 
amplitudes obeying the Ward-Takahashi identities is not renormalizable.

\end{abstract} 

\vskip 10cm
\noindent
PACS: 11.10.Gh; 11.15.-q; 12.20.-m  

\noindent
Keywords: Standard Model, Ward-Takahashi identities, Landau gauge, renormalizability

\newpage

\mysection{Introduction} 

In a recent paper[1], we showed that the standard model is not renormalizable 
in the alpha gauge, the only possible exception being the Landau gauge 
which is a special case of the alpha gauge with alpha equal to zero. In
this paper, we go on to investigate if the standard model in the Landau
gauge is renormalizable.

In the standard model, the number of renormalized or physical parameters
exceed that of independent bare parameters. For example, once we give the
bare vacuum expectation value of the Higgs field $v_0$ a non-zero value, we
generate not just a non-zero renormalized $v$, but also non-zero physical
masses of $W$ and $Z$. Since quantities in the quantum theory of the standard
model have ultraviolet divergences, it is natural to ask if all three
renormalized parameters can be made finite by a choice of one bare
parameter. Similarly, the two bare electro-weak coupling constants
generate renormalized coupling constants of the charged weak current, the
neutral weak current, as well as the electromagnetic current. Can all
these renormalized coupling constants be made finite with proper choices
of the two bare coupling constants ?

For almost three decades, people have agreed that the answer to these
questions is yes. The key point is the validity of the Ward-Takahashi
identities, which, many people believe, render all renormalized quantities
finite. In this paper, we spell out the specific relations satisfied by
the renormalized parameters as consequences of the Ward-Takahashi
identities. 

How do the Ward-Takahashi identities yield relations on the $W$-mass
and the $Z$-mass? Quantum electrodynamics offers a hint. In QED, the
Ward-Takahashi identity for the photon propagator insures that
there are no radiative corrections to the propagator of the longitudinal
photon. This means that the 1PI self-energy for the longitudinal photon
vanishes identically. Now at $k^2 = 0$, the 1PI self-energy for the transverse
photon is equal to that for the longitudinal photon. Consequently, the 1PI
self-energy for the transverse photon vanishes at $k^2 = 0$. Thus the inverse 
of the propagator for the transverse photon vanishes at $k^2 = 0$, signifying 
the existence of a massless vector meson in QED.

Making use of the equality of the 1PI self-energy at zero momentum of the
transverse propagator and that of the longitudinal propagator,
we shall show that the Ward-Takahashi identities indeed lead to relations
among $v$, the renormalized electro-weak coupling constants, and the
physical $W$ mass and the physical $Z$ mass.  These relations resemble their
classical counterparts, with additional factors involving ultraviolet
divergent quantities. 

We shall show that the Ward-Takahashi identities for vertex functions also
lead to specific relations satisfied by the renormalized coupling
constants. There are ultraviolet-divergent quantities in these
relations.

In the derivation of these relations, we shall need the forms of various 
propagators expressed in terms of the 1PI self-energy amplitudes. We
shall list them in the next section.

\mysection{The propagators in the Landau Gauge} 

In the Landau gauge, the unphysical Higgs mesons $\phi^0$ and $\phi^{\pm}$
decouple from
the longitudinal vector mesons and hence we have
$$
\displaystyle 
 G^{\phi^0 \phi^0} (k^2) = \frac{i}{k^2 - \Pi_{\phi^0 \phi^0} (k^2)} \, ,
\eqno(2.1a)
$$ 
and
$$
\displaystyle
  G^{\phi^+ \phi^-} (k^2) = \frac{i}{k^2 - \Pi_{\phi^+ \phi^-} (k^2) } \, ,
\eqno(2.1b)
$$
where $G^{\phi^0 \phi^0}$, for example, is the Fourier transform of 
$ < 0 | T \phi^0 (x) \phi^0 (0) | 0 >$,  
with $T$ the time-ordering operator, and $\Pi_{\phi^0 \phi^0}$ is the 
1PI amplitude of $G^{\phi^0 \phi^0}$.  

The propagators involving longitudinal vector mesons vanish in the Landau
gauge. However, we cannot simply ignore such propagators. This is because,
in the Ward-Takahashi identities, a field operator of a longitudinal vector 
meson is always multiplied by a factor $1/{\alpha}$, where $\alpha$ is the gauge
parameter. Thus, as we take the limit $\alpha \rightarrow 0$, terms in these 
longitudinal propagators which are proportional to $\alpha$ should be retained.
Therefore, we shall need the approximate forms of these propagators up to 
the order of  $\alpha$.  We have  
$$
\displaystyle
 G^{W^+ W^-}_{\mu \nu} (k^2) \approx G^{ZZ}_{\mu \nu} (k^2) \approx
  G^{AA}_{\mu \nu} (k^2) = - i \alpha \frac{k_\mu k_\nu}{ (k^2)^2 } \, ,
\eqno(2.2a)
$$ 
and 
$$
G^{ZA}_{\mu \nu} (k^2) \approx 0 \, ,
\eqno(2.2b)
$$

In (2.2a) and (2.2b), we have chosen the gauge parameter $\alpha$ 
for $A$, $Z$ and $W$
to be the same, and have neglected terms of the order of $\alpha^2$. The
propagators for the transverse components of the vector mesons have not
been included in (2.2). 

The following propagators depend on the specific form of the gauge
fixing terms we use in the effective Lagrangian.  In the main text of 
this paper, we shall choose the gauge fixing terms to be the ones given
by (A.1).  Then we have[1] 
$$
\displaystyle
G^{W^+ \phi^-}_{\mu} (k^2) = G^{\phi^+ W^-}_{\mu} (k^2) \approx
- \frac{i \alpha k_\mu}{M_0} \frac{\Pi_{W^+ \phi^-} (k^2)} {k^2 [k^2
- \Pi_{\phi^+ \phi^-} (k^2)] } \, ,
\eqno(2.3a)
$$ 
and
$$
\displaystyle
G^{V \phi^0}_{\mu} (k^2) = - G^{\phi^0 V}_{\mu} (k^2) \approx 
- \frac{\alpha k_\mu}{M'_0} \frac{\Pi_{V \phi^0} (k^2)} {k^2 [k^2
- \Pi_{\phi^0 \phi^0} (k^2)] } \, ,
\eqno(2.3b)
$$ 
where $V$ is either $A$ or $Z$ and where $M_0$ and $M'_0$  are the 
bare masses of $W$ and $Z$, respectively.

Some of the ghost propagators in the Landau gauge are given by[1] 
$$
\displaystyle
G_{\eta_A \xi_A} (k^2) = \frac{i}{k^2} \frac{1 + g^2_0 \cos^2\theta F(k^2)}{1+g^2_0 F(k^2)} \, ,
\eqno(2.4a)
$$ 
$$
\displaystyle
G_{\eta_A \xi_Z} (k^2) = G_{\eta_Z \xi_A} (k^2) = - \frac{i}{k^2}
 \frac{g^2_0 \cos\theta \sin\theta F(k^2)}{1 + g^2_0 F(k^2)} \, ,
\eqno(2.4b)
$$
and 
$$
\displaystyle
G_{\eta_Z \xi_Z} (k^2) = \frac{i}{k^2} \frac{1+g^2_0 \sin^2\theta F(k^2)}{1+g^2_0 F(k^2)} \, ,
\eqno(2.4c)
$$ 
where 
$$
\Gamma_{\eta_A [W^+_\nu \xi^-]} - \Gamma_{\eta_A [W^-_\nu \xi^+]} 
\equiv k_\nu e_0 F(k^2) \, .  
\eqno(2.5)
$$ 

In (2.4), $\eta_A$ and $\xi_A$ are the ghost fields associated with $A$, $\eta_Z$ 
and $\xi_Z$ are the ghost fields associated with $Z$, $\theta$ 
is the Weinberg angle, and $g_0$ is the bare weak coupling
constant. In (2.5), $\Gamma_{\eta_A [W^+_\nu \xi^-]}$ is the 3-point function 
with the fields $W^+$
and $\xi^-$ joined at the same space-time point and with the propagator of
the external $\eta_A$ omitted.

All of the propagators above have singularities at $k^2 = 0$. We shall
express these propagators in terms of their wavefunction renormalization
constants. We have
$$
\displaystyle 
G_{\phi^a \phi^b} (k^2) \equiv \frac{ i Z_{\phi^a \phi^b} (k^2)}{k^2} \, ,
\eqno(2.6)
$$ 
and
$$
\displaystyle
G_{\eta^i \xi^j} (k^2) \equiv \frac{ i Z_{\eta^i \xi^j} (k^2)}{k^2} \, . 
\eqno(2.7)
$$ 
We shall also put
$$
\displaystyle
G^{W^+ \phi^-}_{\mu} (k^2) \equiv - \frac{ i \alpha k_\mu M_0 
Z_{W^+ \phi^-} (k^2)}{(k^2)^2} \, ,
\eqno(2.8a)
$$ 
and
$$
\displaystyle
G^{V \phi^0}_{\mu} (k^2) \equiv - \frac{\alpha k_\mu M'_0 Z_{V \phi^0}(k^2)}
{(k^2)^2} \, ,
\eqno(2.8b)
$$
where $V$ is either $A$ or $Z$. 
From (2.1) and (2.6) we have
$$
Z_{\phi^0 \phi^0} (k^2) = \bigl[ 1 - \Pi_{\phi^0 \phi^0} (k^2)/k^2 \bigr]^{-1}
\, ,
\eqno(2.9a)
$$ 
and
$$
Z_{\phi^+ \phi^-} (k^2) = \bigl[ 1 - \Pi_{\phi^+ \phi^-} (k^2)/k^2 \bigr]^{-1}
\, .
\eqno(2.9b)
$$ 
From (2.3) and (2.8), we get
$$
\displaystyle
Z_{W^+ \phi^-} (k^2) = \frac{\Pi_{W^+ \phi^-} (k^2)}{M^2_0} 
Z_{\phi^+ \phi^-} (k^2) \, ,
\eqno(2.10a)
$$ 
and
$$
\displaystyle
Z_{V \phi^0} (k^2) = \frac{\Pi_{V \phi^0} (k^2)}{ {M'_0}^2}
Z_{\phi^0 \phi^0} (k^2) \, .
\eqno(2.10b)
$$   
where $V$ is either $A$ or $Z$.
From (2.4), we have
$$
\displaystyle 
Z_{\eta_A \xi_A} (k^2) - \frac{\sin\theta}{\cos\theta} Z_{\eta_A \xi_Z} (k^2) = 1 \, , 
\eqno(2.11a)
$$
and
$$
\displaystyle
Z_{\eta_Z \xi_Z} (k^2) - \frac{\cos\theta}{\sin\theta} Z_{\eta_A \xi_Z} (k^2) = 1 \,  .
\eqno(2.11b)
$$ 

Finally, we turn to the propagators for the transverse components of the
vector mesons. The vector mesons $W$ and $Z$ are massive. Following Peterman,
Stuckelberg[2], Gell-Mann and Low[3], we put the propagator for the
transverse $W$ as
$$
\displaystyle
- \frac{ i Z^T_{W^+W^-} (k^2) }{k^2 - M_W^2} \, ,
\eqno(2.12)
$$  
where $M_W$ is the physical mass of $W$. The ratio of $Z^T_{W^+W^-} (k^2)$ and 
$Z^T_{W^+W^-} (0)$ is
proportional to the effective weak charge[2,3,4].

The propagator for the transverse $Z$ and the transverse $A$ takes a little
more manipulation.  This is because the transverse $Z$ mixes with the
transverse $A$. The mixing matrix of propagators is equal to the inverse
of the matrix
$$
\left[
\begin{array}{cc}
  i ( k^2 - \Pi^T_{AA} ) & - i \Pi^T_{AZ} \\ 
  - i \Pi^T_{AZ} & i ( k^2 - {M'_0}^2 - \Pi^T_{ZZ} ) \\
\end{array}
\right]
\eqno(2.13)
$$ 
where $\Pi^T_{ZZ}$, for example, is the 1PI self-energy amplitude 
for the transverse $Z$.  
Let us diagonalize the matrix in (2.13).  By (4.1b) below, the determinant 
of this matrix vanishes at $k^2 = 0$. Thus so does one of its eigenvalues
at $k^2 = 0$, with the corresponding eigenvector representing the physical
and massless photon which we shall denote as $A'$. The other eigenvector
represents the physical $Z$ meson, which we shall denote as $Z'$. The
particles $A'$ and $Z'$ are related to $A$ and $Z$ by an angle of rotation 
$\Theta (k^2)$.
It is straightforward to find 
$$
\displaystyle
\cot\Theta = \frac{{M'_0}^2 + \Pi_{ZZ} (0)}{\Pi_{AZ} (0)} \, .
\eqno(2.14)
$$ 
where $\Theta = \Theta(0)$.  Let the propagator for the transverse $Z'$ be represented as 
$$
\displaystyle
- \frac{i Z^T_{Z'Z'} (k^2)}{k^2 - M_Z^2} \, ,
\eqno(2.15)
$$
where $M_Z$ is the physical mass of $Z$. The propagator for the transverse 
$A'$ will be represented as 

$$
\displaystyle
- \frac{i Z^T_{A'A'} (k^2)}{k^2} \, .
\eqno(2.16)
$$ 
We have
$$
\displaystyle
 Z^T_{Z'Z'} (0) = \frac{M_Z^2}{{M'_0}^2 + \Pi_{ZZ} (0) + \Pi_{AA} (0) } \, ,
\eqno(2.17)
$$ 
and 
$$
\displaystyle
 Z^T_{A'A'} (0) = {(1 - a)}^{-1} \, ,  
\eqno(2.18)
$$ 
where
$$
\displaystyle
a \equiv \lim_{k^2 \rightarrow 0} \frac{\Pi^T_{AA} (k^2) [{M'_0}^2
+ \Pi^T_{ZZ} (k^2) ] - [\Pi^T_{AZ} (k^2)]^2}
{k^2 [{M'_0}^2 + \Pi_{ZZ} (0) + \Pi_{AA} (0) ] } \, . 
$$

\mysection{The $W$ Mass and the Vacuum Field Value} 

In the Landau gauge, the ghosts directly interact only with the gauge
vector mesons.  According to the Feynman rules, the ghost-ghost-$V$  vertex factor is
proportional to $k^\mu$, where $V$ is a gauge vector meson,
$\mu$ is the polarization of $V$ and  
$k$ is the momentum of the incoming ghost.  Since $\mu$ is always
transverse, $k^\mu$ is equal to $p^\mu$, where $p$ is the momentum 
of the outgoing
ghost. Consequently, this vertex factor vanishes if either the momentum
of the incoming ghost or that of the outgoing ghost vanish. This is a
feature which greatly simplifies some of the Ward-Takahashi identities at
zero momenta discussed below.
          
There are three Ward-Takahashi identities associated with the longitudinal
$W$. We have made use of two of them in a preceding paper and get the
following relation among the 1PI self-energy amplitudes:
$$
\displaystyle
\bigl( 1 + \frac{\Pi_{W^+ W^-} (k^2)}{M^2_0} \bigr)
\bigl( 1 - \frac{\Pi_{\phi^+ \phi^-} (k^2)}{k^2} \bigr) 
= \bigl( 1 + \frac{\Pi_{W^+ \phi^-} (k^2)}{M^2_0} \bigr)^2 \, ,
\eqno(3.1)
$$ 
where $\Pi_{W^+ W^-}$, for example, is the 1PI self-energy amplitude for the
longitudinal $W$. We shall now explore the consequences of the third
identity. 

The bare mass of $W$ is equal to $\displaystyle \frac{1}{2} g_0 v_0$. 
We may choose $v_0$ to be either
the classical vacuum expectation value of the Higgs field at which the
Higgs potential is minimum, or the quantum vacuum expectation value of the
Higgs field. The gauge fixing terms and the ghost terms are different with
these two different choices. Thus the Green functions are different if $v_0$
is chosen differently. However, the physical scattering amplitudes are the
same if physical quantities are gauge invariant.

We shall, in this paper, choose $v_0$ to be the quantum vacuum expectation
value of the Higgs field. With this choice, the third Ward-Takahashi
identities associated with the longitudinal $W$ is, at $k^2 = 0$, 
$$
Z_{W^+ \phi^-} (0) + Z_{\phi^+ \phi^-} (0) = Z_{\eta^+ \xi^-} (0) \, .
\eqno(3.2)
$$ 
We mention that, for $k^2$ not equal to zero, there is a ghost-ghost-Higgs
vertex function appearing in this identity. This vertex function vanishes
at $k^2 = 0$ and does not appear in (3.2), for the reason we mentioned above.

Substituting (2.10a) into (3.2), and making use of (3.1). we reduce (3.2)
into
$$
\displaystyle
\biggl( 1 + \frac{\Pi_{W^+ W^-} (0)}{M^2_0} \biggr) Z_{\phi^+ \phi^-} (0)
= Z^2_{\eta^+ \xi^-} (0) \, .
\eqno(3.3)
$$ 
We will cast (3.3) into a more useful form. We note that the propagator 
for the transverse W may be expressed by its 1PI amplitude as:
$$
\displaystyle
- \frac{i}{k^2 - M^2_0 - \Pi^T_{W^+ W^-} (k^2)} \, ,
\eqno(3.4)
$$ 
where $\Pi^T_{W^+ W^-}$ is the 1PI self-energy amplitude for the 
transverse $W$. From
(2.12) and (3.4), we get
$$
\displaystyle
 Z^T_{W^+ W^-} (0) = \frac{M^2_W}{M^2_0 + \Pi^T_{W^+ W^-} (0)} \, .
\eqno(3.5)
$$ 
Making use of the fact that $\Pi^T_{W^+ W^-} (0)$ and $\Pi_{W^+ W^-} (0)$ 
are equal, we reduce (3.3) into 
$$
\displaystyle
M_W = \frac{1}{2} g_0 v_0 Z_{\eta^+ \xi^-} (0) \sqrt{Z^T_{W^+ W^-} (0) /
Z_{\phi^+ \phi^-}(0) } \, .
\eqno(3.6)
$$  
Finally, $v$, the renormalized vacuum expectation value of the Higgs field,
is equal to $v_0$ divided by $\sqrt{Z_H(0)}$, where $Z_H (0)$ is the wavefunction
renormalization constant for the physical Higgs field $H$. Thus we have
$$
\displaystyle
  v = \frac{v_0}{\sqrt{Z_H(0)}} \, .
\eqno(3.7)
$$ 
Hence (3.6) becomes 
$$
\displaystyle
\frac{M_W}{v} = \frac{1}{2} g_0 Z_{\eta^+ \xi^-} (0)
\sqrt{Z^T_{W^+ W^-} (0)} \sqrt{\frac{Z_H(0)}{Z_{\phi^+ \phi^-} (0)}}  \, .
\eqno(3.8)
$$ 
If the standard model is renormalizable, the right-side of (3.8) must be
finite.

\noindent
We close this section with two comments:

\noindent
(a)The relationship (3.8) holds in the Landau gauge, a special case of
the alpha gauge. One may derive its counterpart in a general alpha gauge. 

\noindent
(b)The $W$ meson is unstable, thus the propagator for the transverse W does 
not have a pole at a real value of $k^2$.  We may define $M$ to be the value
of $k^2$ at which the real part of the inverse of the propagator for the
transverse $W$ vanishes. This will provide a subtraction condition for the
propagator of the transverse $W$.

\mysection{The $Z$ Mass and the Vacuum Field Value} 

Next we consider the Ward-Takahashi identities  associated with the
longitudinal $A$ and the longitudinal $Z$.  There are nine of them in all.
We have extracted the consequences from seven of them. Among others, we
have
$$
\displaystyle
\biggl[ 1 + \frac{\Pi_{ZZ} (k^2)}{{M'_0}^2} \biggr] \biggl[ 1 - 
\frac{\Pi_{\phi^0 \phi^0} (k^2)}{k^2} \biggr] = \biggl[ 1 + 
\frac{\Pi_{Z \phi} (k^2)}{{M'_0}^2} \biggr]^2 \, ,
\eqno(4.1a)
$$ 
$$
\Pi_{AA} (k^2) [{M'_0}^2 + \Pi_{ZZ} (k^2) ] = [\Pi_{AZ} (k^2)]^2 \, .
\eqno(4.1b)
$$ 
and 
$$
\displaystyle
\Pi_{AA} (k^2) \biggl[ 1 - \frac{\Pi_{\phi^0 \phi^0} (k^2)}{k^2} \biggr]
= \frac{[\Pi_{A \phi} (k^2)]^2}{{M'_0}^2} \, ,
\eqno(4.1c)
$$ 
where $\Pi_{ZZ}$, for example, is the 1PI self-energy amplitude for the
longitudinal $Z$.  The remaining two Ward-Takahashi identities are, 
at $k^2 = 0$,
$$
Z_{A \phi} (0) = Z_{\eta_A \xi_Z} (0) \, ,
\eqno(4.2a)
$$
and  
$$
\displaystyle
\bigl[ 1 + \frac{\Pi_{Z \phi} (0)}{{M'_0}^2} \bigr] Z_{\phi^0 \phi^0} (0)
= Z_{\eta_Z \xi_Z} (0) \, .
\eqno(4.2b)
$$ 
From (4.1a) and (4.2b), we get
$$
\displaystyle
\bigl[ 1 + \frac{\Pi_{ZZ} (0)}{{M'_0}^2} \bigr] Z_{\phi^0 \phi^0} (0)
= [ Z_{\eta_Z \xi_Z} (0) ]^2 \, .
\eqno(4.3)
$$ 
Making use of eq. (2.17), we get
$$
\displaystyle
\frac{M_Z}{v} = \frac{1}{2} \frac{g_0}{\cos\theta} \sqrt{Z^T_{Z Z'}(0)}
\sqrt{\frac{Z^2_{\eta_Z \xi_Z}(0)}{Z_{\phi^0 \phi^0} (0)} +
\frac{\Pi_{AA}(0)}{{M'_0}^2} } \sqrt{Z_H(0)} \, .
\eqno(4.4)
$$ 
The right-side of (4.4) must be finite if the standard model is
renormalizable.

\mysection{Identities for Three Point Functions} 

In this section we study the consequences of the Ward-Takahashi identities
on three-point functions in the standard model.
We shall begin by summarizing a number of relations among the wavefunction
renormalization constants which will be used in this section.

From (4.2a) and (4.2b), we obtain
$$
\displaystyle
\frac{Z_{\eta_A \xi_Z} (0)}{Z_{\eta_Z \xi_Z} (0)} 
= \tan\Theta \, ,
\eqno(5.1)
$$ 
where we have made use of (2.10b), (2.14) and (4.1). Together with 
(2.11a), and (2.11b), (5.1) give 
$$
\displaystyle
Z_{\eta_Z \xi_Z} (0) = \frac{\sin\theta\, \cos\Theta}{\sin(\theta-\Theta)} \, ,  
\eqno(5.2a)
$$ 
$$
\displaystyle
Z_{\eta_A \xi_Z} (0) = \frac{\sin\theta\, \sin\Theta}{\sin(\theta-\Theta)} \, ,
\eqno(5.2b)
$$ 
and
$$
\displaystyle
Z_{\eta_A \xi_A} (0) = \displaystyle \frac{\frac{1}{2}\sin(2\theta)\, 
\cos\Theta - \cos(2\theta)\, \sin\Theta}
{\cos\theta\, \sin(\theta -\Theta)} \, .
\eqno(5.2c)
$$ 
Similarly, we may derive from (2.10b) that
$$
\displaystyle
\frac{Z_{A \phi^0} (0)}{Z_{Z \phi^0} (0) + Z_{\phi^0 \phi^0} (0)}
= \tan\Theta \, .
\eqno(5.3)
$$ 
It is also straightforward to find that
$$
\displaystyle 
[ Z_{A \phi^0} (0)  ]^2 + [Z_{Z \phi^0} (0) + Z_{\phi^0 \phi^0} (0) ]^2
= \biggl[ 1 + \frac{\Pi_{AA} (0) + \Pi_{ZZ} (0)}{{M'_0}^2} \biggr]
Z_{\phi^0 \phi^0} (0) \, .
$$
Thus, by (2.17), we have
$$
\displaystyle
 Z_{A \phi^0} (0) = \frac{M_Z}{M'_0} \sqrt{\frac{Z_{\phi^0 \phi^0}}
{Z^T_{Z' Z'} (0)} } \sin\Theta \, ,
\eqno(5.4a)
$$ 
and 
$$
\displaystyle
 Z_{Z \phi^0} (0) + Z_{\phi^0 \phi^0} (0) = \frac{M_Z}{M'_0}
\sqrt{\frac{Z_{\phi^0 \phi^0} (0)}{Z^T_{Z' Z'}(0)} } \cos\Theta \, .
\eqno(5.4b)
$$ 
By (4.1), eq.(2.14) can also be written as
$$
\displaystyle
\cos\Theta = \sqrt{\frac{{M'_0}^2 + \Pi_{ZZ} (0)}{{M'_0}^2 + \Pi_{ZZ} (0)
 + \Pi_{AA} (0)} } \, ,
\eqno(5.5a)
$$  
and
$$
\displaystyle
\sin\Theta = \sqrt{\frac{\Pi_{AA} (0) }{{M'_0}^2 + \Pi_{ZZ} (0) + \Pi_{AA} (0) } } \, .
\eqno(5.5b)
$$ 
We shall now derive the Ward identities for the interaction of a lepton $l$
with a gauge meson. 

The first such identity is obtained by setting to zero the vacuum expectation 
value of the BRST variation of   
$$
  T i \eta_A (x) l(y) \bar{l}(z) \, .
$$ 
We get
$$
\begin{array}{ll}
\displaystyle
 \frac{1}{\alpha} < 0 | T \partial_\mu A^\mu(x) l(y) \bar{l}(z) | 0 > & 
 =   \displaystyle i g_0 < 0 | T i \eta_A (x) \biggl[\frac{\frac{1}{2}L - \sin^2\theta}{\cos\theta} \xi_Z (y) 
+ \sin\theta \xi_A (y) \biggr] l(y) \bar{l} (z) | 0> \\
& \displaystyle - \frac{i g_0}{\sqrt{2}} < 0 | T i \eta_A (x) \xi^-(y) \nu(y) \bar{l}(z)
| 0 > \\
& \displaystyle - i g_0 < 0 | T i \eta_A(x) l(y) \bar{l}(z) 
\biggl[\frac{\frac{1}{2} R 
- \sin^2\theta}{\cos\theta} \xi_Z(z) + \sin\theta \xi_A(z) \biggr] | 0> \\
& \displaystyle +\frac{i g_0}{\sqrt{2}} < 0 | T i \eta_A(x) l(y) \bar{\nu}(z) \xi^+(z)
| 0 > \\ 
\end{array}
\eqno(5.6)
$$ 
where $\alpha$ is the gauge parameter, $\nu$ is the neutrino associated 
with $l$, $\displaystyle L = \frac{1}{2} (1 + \gamma_5)$ and 
$\displaystyle R = \frac{1}{2} (1 - \gamma_5)$.

Next we set to zero the vacuum expectation value of the BRST variation of
$$
  T i \eta_Z(x) l(y) \bar{l}(z) \, .
$$ 
We get
$$
\begin{array}{ll}
\displaystyle
 < 0 | T \biggl[ \frac{\partial_\mu Z^\mu(z)}{\alpha} + M'_0 \phi^0(x) 
\biggr] l(y) \bar{l}(z) | 0 > \\ 
=  \displaystyle i g_0 < 0 | T i \eta_Z(x) \biggl[\frac{\frac{1}{2}L - \sin^2\theta}{\cos\theta} \xi_Z(y)
+ \sin\theta \xi_A(y) \biggr] l(y) \bar{l}(z) | 0 > \\
{\hspace{1em}} \displaystyle - \frac{i g_0}{\sqrt{2}} < 0 | T i \eta_Z(x) \xi^-(y) \nu(y) \bar{l}(z)
| 0 > \\
{\hspace{1em}} \displaystyle - i g_0 < 0 | T i \eta_Z(x) l(y) \bar{l}(z) 
\biggl[ \frac{\frac{1}{2}R - 
\sin^2\theta}{\cos\theta} \xi_Z(z) + \sin\theta \xi_A(z) \biggr] | 0 > \\
{\hspace{1em}} \displaystyle + \frac{i g_0}{\sqrt{2}} < 0 | T i \eta_Z(x) l(y) \bar{\nu}(z)
\xi^+(z) | 0 > \, .
\end{array} 
\eqno(5.7)
$$ 

We note that the longitudinal $A$ and the longitudinal $Z$ mix with the
unphysical neutral Higgs meson $\phi^0$. Thus the external longitudinal
photon in the left-side of (5.6), for example, may propagate into a
photon, or a $\phi^0$. For this reason, the term on the left-side
of (5.6) gives rise to two terms, each of which corresponds to a channel
of propagation. Similarly, the term on the left-side of (5.7) also give
rise to two terms. We mention that, by (2.2b),the longitudinal A does not
propagate into the longitudinal $Z$ in the Landau gauge.

We shall take the Fourier transform of (5.6) and (5.7),denoting the
momenta of the outgoing lepton, the incoming lepton, and the outgoing
gauge meson by $p'$, $p$, and $k$, respectively, with
$$
                     p = p' + k.
$$
We take $\alpha$ to zero to get to the Landau gauge. Then we multiply the
resulting expression by $i k^2$ (to get rid of the propagator of the
external longitudinal photon) as well as by  $S_l^{-1} (p)$ from the right 
and  $S_l^{-1} (p')$ 
from the left (to eliminate the propagators of the external leptons). We
then differentiate the resulting equation with respect to $k_\mu$ with $p$
fixed and on-shell, take the limit $k \rightarrow 0$ and insert the equation 
between physical lepton spinor functions. We get
$$
\begin{array}{ll}
\displaystyle
- i \Gamma^\mu_{l \bar{l} A} (p,p,0) - M'_0 Z_{A \phi^0} (0)
\frac{\partial}{\partial k_\mu} \Gamma_{l \bar{l} \phi^0} (p-k,p,k)
|_{k=0} \\
{\hspace{6em}} =  \displaystyle \frac{g_0 Z_{\eta_A \xi_Z} (0)}{Z_l (m^2)} \gamma^\mu
\frac{\frac{1}{2}L - \sin^2\theta}{\cos\theta} + \frac{e_0 Z_{\eta_A \xi_A} (0)}{Z_l (m^2)} 
\gamma^\mu \, ,
\end{array}
\eqno(5.8)
$$ 
and
$$
\begin{array}{ll}
\displaystyle
- i \Gamma^\mu_{l \bar{l} Z} (p,p,0) - M'_0 [Z_{Z \phi^0} (0) + 
Z_{\phi^0 \phi^0} (0)] \frac{\partial}{\partial k_\mu} 
\Gamma_{l \bar{l} \phi^0} (p-k,p,k) |_{k=0} \\
{\hspace{5em}} = \displaystyle \frac{g_0 Z_{\eta_Z \xi_Z} (0)}{Z_l (m^2)} \gamma^\mu
\frac{\frac{1}{2} L - \sin^2\theta}{\cos\theta} + e_0 \frac{Z_{\eta_Z \xi_A} (0)}{Z_l (m^2)}
\gamma^\mu \, ,
\end{array}
\eqno(5.9)
$$ 
where $Z_l (p^2)$ is the wavefunction renormalization constant for the lepton,
$p^2 = m^2$, and $m$ is the mass of the lepton. There are 4-point amplitudes in
the Ward-Takahashi identities, but they vanish as we set $k = 0$. 

Since $A'$, not $A$, is  the physical photon field, we take the difference of
(5.8) multiplied by $\cos\Theta$ and (5.9) multiplied by $\sin\Theta$, getting  
$$
\displaystyle
- i \Gamma^\mu_{l \bar{l} A'} (p,p,0) = e_0 \frac{Z_{\eta_A \xi_A}(0)\cos\Theta 
- Z_{\eta_Z \xi_A} (0) \sin\Theta}{Z_l (m^2)} \gamma^\mu \, .
\eqno(5.10)
$$ 
The left-side of (5.10) multiplied by $Z_l(m^2) \sqrt{Z^T_{A'A'}(0)}$
 is equal to
$$
   e \gamma^\mu \, ,
\eqno(5.11)
$$ 
where $e$ is the renormalized electric charge. Thus we have
$$
\displaystyle
  e = e_0 \sqrt{Z^T_{A'A'} (0)} \frac{\cos(\theta - \Theta)}{\cos \theta} \, ,
\eqno(5.12)
$$ 
where (5.2) has been used. 

The renormalized charge given by (5.12) is universal, i.e., the same for
all leptons[5]. We note that the form in (5.12) differs from its
counterpart in QED by the last factor in (5.12).  In order that the
standard model is renormalizable, the right-side of (5.12) is required to
be finite.

Taking the sum of (5.8) multiplied by $\sin\Theta$ and (5.9) multiplied by 
$\cos\Theta$,
we get
$$
\begin{array}{ll}
& \displaystyle - i \Gamma^\mu_{l \bar{l} Z'} (p,p,0) - M_Z \sqrt{\frac{Z_{\phi^0 \phi^0}}
{Z^T_{Z'Z'}(0)}} \frac{\partial}{\partial k_\mu} \Gamma_{l \bar{l} \phi^0}
(p-k,p,k) |_{k=0} \\
= & \displaystyle \frac{\sin \theta}{\sin(\theta - \Theta)} \frac{g_0}{Z_l(m^2)}
\gamma^\mu \frac{\frac{1}{2}L - \sin^2\theta}{\cos\theta} + \frac{\sin\Theta \sin(2 \theta
- \Theta)}{\cos\theta \sin(\theta - \Theta)} \frac{e_0}{Z_l(m^2)}
\gamma^\mu \, ,
\end{array} 
\eqno(5.13)
$$
where $p^2 = m^2$.

Next we discuss the neutral current for the neutrino. The counterparts of
(5.8) and (5.9) are
$$
\displaystyle
- i \Gamma^\mu_{\nu \bar{\nu} A} (p,p,0) - M'_0 Z_{A \phi^0} (0)
\frac{\partial}{\partial k_\mu} \Gamma_{\nu \bar{\nu} \phi^0} (p-k,p,k) 
|_{k=0} = - \frac{g_0 Z_{\eta_A \xi_Z} (0)}{2 \cos\theta Z_{\nu} (0)} \gamma^\mu \, , 
\eqno(5.14)
$$  
and
$$
\displaystyle
- i \Gamma^\mu_{\nu \bar{\nu} Z} (p,p,0) - M'_0 [Z_{Z \phi^0} (0)
+ Z_{\phi^0 \phi^0} (0) ] \frac{\partial}{\partial k_\mu}
\Gamma_{\nu \bar{\nu} \phi^0} (p-k,p,k) |_{k=0}
= - \frac{g_0 Z_{\eta_Z \xi_Z} (0)}{2 \cos\theta Z_{\nu} (0) } \gamma^\mu \, ,
\eqno(5.15)
$$ 
where $p^2 = 0$.  Thus the counterpart of (5.10) is
$$
\Gamma^{\mu}_{\nu \bar{\nu} A'} (p,p,0) = 0 \, ,
\eqno(5.16)
$$ 
which says that the renormalized charge of the neutrino is rigorously 
zero. The counterpart of (5.13) is
$$
\begin{array}{ll}
\displaystyle
- i \Gamma^\mu_{\nu \bar{\nu} Z'} (p,p,0) - M_Z \sqrt{\frac{Z_{\phi^0 \phi^0}
(0)}{Z^T_{Z'Z'}(0)}} \frac{\partial}{\partial k_\mu} 
\Gamma_{\nu \bar{\nu} \phi^0} (p-k,p,k) |_{k=0} \\
{\hspace{6em}} = \displaystyle - \frac{\tan \theta}{\sin(\theta - \Theta)}
 \frac{g_0}{2 Z_\nu (0)} \gamma^\mu \, ,
\end{array}
\eqno(5.17)
$$  
where $p^2 = 0$.  

Let us multiply (5.17) by $\sqrt{Z^T_{Z'Z'}(0)} Z_\nu(0)$. Then the first and the second 
term on the left-side of the resulting equation are proportional to the
renormalized $\nu-\bar{\nu}-Z'$ vertex function and the renormalized
$\nu - \bar{\nu} - \phi^0$  
vertex function, respectively. If these two renormalized vertex functions
are finite, so must be the right-side of the resulting equation. Thus we
require that
$$
\displaystyle
 g_Z \equiv \frac{\tan \theta}{\sin(\theta - \Theta)} \frac{g_0}{2}
  \sqrt{Z^T_{Z'Z'}(0)} \, ,
\eqno(5.18)
$$
to be finite.
 
Finally, we consider the charged weak current. We set to zero the vacuum
expectation value of the BSRT variation of   
$$
    T i \eta^-(x) \nu(y) \bar{l}(z) \, .
$$
We get
$$
\begin{array}{ll}
\displaystyle
< 0 | T (\frac{\partial^\mu W^-_\mu(x)}{\alpha} - i M_0 \phi^-(x)) \nu(y)
\bar{l}(z) | 0 > \\
 =  \displaystyle - i g_0 < 0 | T i \eta^-(x) \biggl[ \frac{\xi_Z(y) \nu(y)}
{2 \cos\theta} + \frac{\xi^+(y) L\, l(y)}{\sqrt{2}} \biggr] \bar{l}(z) | 0 > \\
{\hspace{1em}} \displaystyle - i g_0 < 0 | T i \eta^-(x) \nu(y) \biggl[ \frac{\frac{1}{2}
R - \sin^2\theta}{\cos\theta} \xi_Z(z) + \sin\theta \xi_A(z) \biggr] \bar{l}(z) | 0 > \\
{\hspace{1em}} \displaystyle + \frac{i g_0}{\sqrt{2}} < 0 | T i \eta^-(x) \nu(y)
\bar{\nu}(z) \xi^+(z) | 0 > \, .
\end{array}
\eqno(5.19)
$$  
As before, we take the limit $\alpha$ going to zero and take the Fourier
transform of eq.(5.19), with the momentum  of the outgoing neutrino
denoted by $(p-k)$ and that of the incoming lepton denoted by $p$. We
multiply the Fourier transform of (5.19) by $i k^2$ as well as by 
$S_\nu^{-1} (p-k)$ from the left and $S_l^{-1} (p)$ from the right, 
differentiate  with respect to $k$
with $p$ fixed, and set $k$ to zero.  Since the masses of the neutrino and the
lepton are different, we cannot make both of these particles to be on the
mass-shell. We shall choose to have the electron on the mass-shell,
i.e., we choose $p^2 = m^2$.  We apply the resulting expression on the lepton
spinor function, setting $\pslash$  operating on the lepton spinor function to
equal to $m$. We get
$$
\begin{array}{ll}
\displaystyle 
 - i \Gamma^\mu_{\nu \bar{l} W^-} (p,p,0) + i M_W \sqrt{\frac{Z_{\phi^+
\phi^-} (0)}{Z^T_{W^+ W^-} (0)} } \frac{\partial}{\partial k_\mu}
\Gamma_{\nu \bar{l} \phi^-} (p-k,p,k) |_{k=0} \\
\displaystyle =  - \frac{g_0}{\sqrt{2}} \frac{Z_{\eta^- \xi^+} (0)}
{Z_\nu (m^2)} \gamma^\mu L \, ,
\end{array}
\eqno(5.20)
$$  
where $p^2 = m^2$, and where (3.5) has been used.

Let us multiply equation (5.20) by $\sqrt{Z^T_{W^+ W^-}(0) Z_l(m^2) 
Z_\nu(m^2)}$, then the first term
and the second term in the left-side of the resulting equation are
proportional to the renormalized $\nu-l-W$ vertex the renormalized
$\nu-\bar{l}-\phi^-$  
vertex, respectively. If these two renormalized vertices are finite, so
must be the right-side of the resulting equation. Thus
$$
\displaystyle 
 g_W \equiv g_0 Z_{\eta^+ \xi^-}(0) \sqrt{\frac{Z_l(m^2)}{Z_\nu(m^2)}
 Z^T_{W^+ W^-}(0)} \, ,
\eqno(5.21)
$$ 
is required to be finite.

\mysection{Discussion} 

As theoretical physicsists all know, the forms of the Ward-Takahashi
identities in QED are relatively simple. People also know that
these identities have profound consequences, two of them being the
vanishing of the photon mass and the universality of the electric charge.
All of these consequences are easily extracted from the Ward-Takahashi
identities in QED.

In contrast, the Ward-Takahashi identities in Yang-Mills theories are
notoriously complex in form. There are numerous terms in these identities 
due to the existence of interacting ghosts. Such complexities mar the
rigorous implications of these identities which are more difficult to
explore.

As an example, consider using these identities in the standard model in
the Feynman gauge and investigate the question of the universality of the 
electric charge. The ratio of $Z_{e \bar{e} A}$ and $Z_e$ obtained 
from these identities
is not identically unity---the value of this ratio in QED. Instead, it is
equal to a sum of amplitudes.  It is possible to calculate this sum
perturbatively and showed that, to the one-loop order, it is independent
of the lepton mass.  But to prove this true to all orders on the basis of
these identities in the Feynman gauge appears difficult.

But these identities do give simple and exact consequences. For example,
it is easy to prove that the mass of the photon is strictly zero with the
Ward-Takahashi identities. It is also possible to prove charge universality 
from these identities if one uses the unitary gauge[5].

Additional consequences from these identities are found if one uses the
Landau gauge. This is because these identities at $k^2 = 0$ simplify in the
Landau gauge. One finds relationships between the $W$-mass and the $Z$-mass
with the renormalized vacuum expectation value of the Higgs field:
$$
\displaystyle
 \frac{M_W}{\frac{1}{2} v g_W} = \sqrt{\frac{Z_H(0)}{Z_{\phi^+ \phi^-}(0)}
 \frac{Z_\nu(m^2)}{Z_l(m^2)}}  
\eqno(6.1)
$$ 
and 
$$
\displaystyle
\frac{M_Z}{v g_Z} = \frac{\sin(\theta - \Theta)}{\sin \theta}
\sqrt{\biggl[ \frac{Z^2_{\eta_Z \xi_Z} (0)}{Z_{\phi^0 \phi^0} (0)}
+ \frac{\Pi_{AA}(0)}{{M'_0}^2} \biggr] Z_H(0)} \, .
\eqno(6.2)
$$  
These are non-perturbative and quantum mechanical expressions for the
masses of $W$ and $Z$ generated by spontaneous symmetry breaking. If one
neglects quantum corrections and set the right-sides of (6.1) and (6.2)
to unity, they are reduced to the well-known classical formulae for
the $W$ mass and the $Z$ mass.

One also finds the following three renormalized electro-weak coupling
constants:
$$
\displaystyle
e \equiv e_0 \sqrt{Z^T_{A'A'}(0)} \frac{\cos(\theta - \Theta)}{\cos \theta} \, ,
\eqno(6.3)
$$
$$
\displaystyle
g_Z \equiv \frac{g_0}{2} \sqrt{Z^T_{Z'Z'}(0)} \frac{\tan \theta}{\sin(\theta
-\Theta)} \, ,
\eqno(6.4)
$$ 
and 
$$
\displaystyle
g_W \equiv g_0 \sqrt{Z^T_{W^+ W^-}(0)} \sqrt{\frac{Z_l(m^2)}{Z_\nu(m^2)}}
Z_{\eta^+ \xi^-}(0) \, .
\eqno(6.5)
$$  
Equation (6.3) gives the electric charge $e$ in the standard model. 
Equation (6.4) gives the coupling constant $g_Z$ for the neutral weak
current.  And (6.5) gives the coupling constant $g_W$ for the charged weak
current. 

In addition, by multiplying (5.13) with $Z_l(m^2) \sqrt{Z^T_{Z'Z'}(0)}$, we get
$$
\begin{array}{ll}
\displaystyle
- i Z_l(m^2) \sqrt{Z^T_{Z'Z'}(0)} \Gamma^\mu_{l \bar{l} Z'}(p,p,0) 
- M_Z Z_l (m^2) \sqrt{Z_{\phi^0 \phi^0}(0)} \frac{\partial}{\partial k_\mu}
\Gamma_{l \bar{l} \phi^0} (p-k,p,k) |_{k=0} \\
 = g_Z \gamma^\mu [ L - 2 \sin^2(\theta-\Theta) ] \, ,
\end{array}
\eqno(6.6)
$$  
where $p^2 = m^2$. We note that both terms in the left-side of (6.6) are
renormalized amplitudes. Thus the right-side of (6.6) must be
ultraviolet finite if the standard model is renormalizable. 

While the Ward-Takahashi identities in the standard model do lead to 
relationships among various renormalized quantities, they by no means 
imply that all of these quantities can be chosen ultraviolet finite.
This is because there are ultraviolet divergent quantities in these
relations. For the renormalized quantities in the above equations to be
finite, the right sides of these equations must be finite. In particular,
if $g_Z$ and the right-side of(6.6) are both finite, $\theta - \Theta$ must be
ultraviolet finite for the standard model to be renormalizable.

From (6.3) and (6.4), we get
$$
\displaystyle
\frac{g_Z}{e} \sin[2 (\theta - \Theta) ] = \sqrt{\frac{Z^T_{Z'Z'}(0)}
{Z^T_{A'A'}(0)}} \, .
\eqno(6.7)
$$ 
Thus the right-side of (6.7) must be ultraviolet finite for the standard
model to be renormalizable.
But the right-side of (6.7) is not ultraviolet finite. Indeed, this ratio
has been calculated up to one loop and is known to be ultraviolet 
divergent[5].  

In conclusion, a quantum gauge field theory is not completely predictive
without a prescription of how the ultraviolet divergences are handled. As
is well-known, the prescription in QED is that the divergent amplitudes
obey the Ward-Takahashi identities. This leads to predictions in spectacular 
agreements with experiments. If one uses the same prescription for the
quantum theory of the standard model, one also finds exact results such as
charge universality, the vanishing of the electric charge of the neutrino, 
and the vanishing of the photon mass. This is borne out by experiments to
an extremely high degree, as the sum  of the electron charge and the proton  
charge is less than $10^{-21} e$, the neutrino charge is less than $10^{-13}e$,
and the photon mass is less than $6 \times 10^{-16} eV$.  On the other hand, this prescription
in the standard model leads to a non-renormalizable theory, as not all
renormalized quantites can be chosen finite. We believe that the foundation 
of the quantum theory of the standard model remains to be laid.

\newpage
\begin{center}
\large {\bf Appendix}
\end{center}

In the main text of this paper, the gauge fixing terms of the Lagrangian
are  
$$
\displaystyle
- \frac{1}{\alpha} (\partial^\mu W^+_\mu + i \alpha M_0 \phi^+ ) 
(\partial^\nu W^-_\nu - i \alpha M_0 \phi^- ) - \frac{1}{2 \alpha}
(\partial_\mu Z^\mu + \alpha M_0 \phi^0 )^2 - \frac{1}{2 \alpha} 
(\partial_\mu A^\mu)^2 \, .
\eqno(A.1)
$$ 
In this Appendix,we discuss briefly how the formulae are modified if the
gauge fixing terms are chosen to be, instead,
$$
\displaystyle
- \frac{1}{\alpha} (\partial^\mu W^+_\mu)(\partial^\nu W^-_\nu)
- \frac{1}{2 \alpha} (\partial_\mu Z^\mu)^2 - \frac{1}{2 \alpha} 
(\partial_\mu A^\mu)^2 \, .
\eqno(A.2)
$$ 
With the gauge fixing terms given by (A2), all the propagators in the 
limit $\alpha \rightarrow 0$  are of the same form as the ones in Sec.2 except the ones
given by (2.3a) and (2.3b), which are replaced by              
$$
G^{W^+\phi^-}_\mu (k) = G^{\phi^+W^-}_\mu (k) \approx \displaystyle 
- \frac{i \alpha k^\mu}{M_0} \frac{M^2_0 + \Pi_{W\phi}}{(k^2)^2 
(1 - \frac{\Pi_{\phi^+\phi^-}}{k^2}) } \, ,
\eqno(A.3a)
$$ 
and 
$$
\displaystyle
G^{Z\phi^0}_\mu = - G^{\phi^0 Z}_\mu \approx - \frac{\alpha k^\mu}{M'_0}
\frac{M'^2_0 + \Pi_{Z \phi^0}}{k^2 [k^2 - \Pi_{\phi^0\phi^0}(k^2)]} \, .
\eqno(A.3b)
$$  
Thus all wavefunction renormalization constants remain the same as
before except the ones given by (2.10a) and (2.10b), which  are replaced
by
$$
\displaystyle
 Z_{W^+\phi^-} (k^2) = ( 1 + \frac{\Pi_{W\phi}(k^2)}{M^2_0})
  Z_{\phi^+\phi^-}(k^2) \, ,
\eqno(A.4a)
$$ 
and
$$
\displaystyle
Z_{Z\phi^0} (k^2) = ( 1 + \frac{\Pi_{Z\phi}(k^2)}{M'^2_0} )
 Z_{\phi^0\phi^0} (k^2) \, .
\eqno(A.4b)
$$ 
The relations among the 1PI self-energy amplitudes given by (3.1) and
(4.1) remain valid.

The Ward-Takahashi identity (3.2) is replaced by 
$$
Z_{W^+\phi^-} (0) = Z_{\eta^+\xi^-}(0) \, .
\eqno(A.5)
$$ 
while the Ward identities (4.2a) and (4.2b) remain the same as before.
Because of these changes, some of the intermediate formulae are now
different. For example, we have, instead of (5.3),
$$
\displaystyle
\frac{Z_{A\phi^0}(0)}{Z_{Z\phi^0}(0)} = \tan\Theta \, .
\eqno(A.6)
$$ 
However, the final formulae (6.1)--(6.6) stay the same.

\newpage
\begin{center}
\large \bf {References}
\end{center}

\begin{enumerate}

\item{H. Cheng and S.P. Li, The Standard Model In The Alpha Gauge Is Not
Renormalizable (unpublished).}

\item{E.C.G. Stuckelberg and A.Petermann, Helv. Phys. Acta 26,499 (1953).}

\item{M.Gell-Mann and F.E.Low, Phys.Rev.95,1300 (1954).}

\item{L.Baulieu and R. Coquereaux, Ann. of Phys. 140,163 (1982).}

\item{K.Aoki, Z. Hioki, R.Kawabe, M. Konuma, and T. Muta, Prog. Theor.Phys.
Suppl. 73, 1 (1982).}

\end{enumerate}

\end{document}